\documentclass{aa}
\usepackage{graphics}
\usepackage{astron}
\usepackage{amssymb}
\usepackage{dcolumn}

\begin{document}

	\thesaurus{20(11.06.2; 11.07.1; 11.12.2; 11.16.1; 11.09.5; 11.19.6)}
%
	\title{Structure and stellar content of dwarf galaxies}
 	\thanks{Based on observations made at Observatoire de Haute
 	Provence (CNRS), France} \subtitle{IV. B and R
 	photometry of dwarf galaxies in the CVnI cloud.}
	
	\titlerunning{Structure and stellar content of dwarf
	galaxies. IV}
	\authorrunning{Bremnes et al.}
	
	\author{T. Bremnes \inst{1}
	\and B. Binggeli \inst{1}
	\and P. Prugniel \inst{2}}

	\institute{Astronomical Institute, University of Basel,
		Venusstrasse 7, CH-4102 Binningen, Switzerland
		\and Observatoire de Lyon, 
		F-64561 St.\ Genis-Laval Cedex, France}

	\offprints{T. Bremnes}
	\mail{bremnes@astro.unibas.ch}

\date{Received date / Accepted date}

\maketitle

\begin{abstract}
We have carried out CCD photometry in the Cousins $B$ and $R$ bands of
15 galaxies in the Canes Venatici I cloud. Total magnitudes, effective
radii, effective surface brightnesses, as well as galaxy radii at
various isophotal levels in both colors were determined.  Best-fitting
exponential parameters and color gradients are also given for these
galaxies. The photometric parameters presented here will analyzed in a
forthcoming paper, together with previously published data for nearby
dwarf galaxies.

\end{abstract}

\section{Introduction}
Continuing a series of papers on dwarf galaxies within 10 Mpc
\cite{1998A&AS..129..313B,les99,1999A&AS..137..337B}, hereafter Papers
I, II and III, we here present photometric data on dwarf galaxies in
the Canes Venatici I cloud (CVnI). The cloud is composed of some 40
galaxies distributed in the region of the sky delimited by $ 11\fh5
<\alpha< 14\fh0$, $ +20\degr <\delta< +60\degr$
\cite{1998A&AS..128..459M} and has a mean heliocentric radial velocity
of $\mathopen{<}v_{\mathrm{hel}}\mathclose{>} \approx 300
\mathrm{km\,s^{-1}}$.  A map showing the distribution of these objects
on the sky is shown by Makarova et al.~\cite*{1998A&AS..128..459M} in
their Fig.~1.  The structure of the cloud is somewhat different from
the groups studied in Papers I,II and III in that there is no dominant
galaxy, like in the M81 or M101 groups. It is constituted of several
loose aggregations of galaxies, mostly of late types. This study
therefore complements Papers I,II and III towards globally lower
density environments.  Accurate distances are becoming available as
photometric distance determinations\footnote{radial velocities are not
reliable distance indicators for nearby galaxies because of possibly
large peculiar velocities.}
\cite{1997AstL...23..378M,1997AstL...23..514G,1998A&AS..128..325T,1998A&AS..128..459M,1998A&AS..131....1K}.
The combination of distance information and photometry will enable a
thorough study of the photometric parameters of these systems.

In the following section we describe the sample and the imaging, in
Sect.\ \ref{reductions} we describe the reduction procedure. The
results are presented in Sect. \ref{results}. Model-free parameters,
i.e. total magnitudes, effective radii and effective surface
brightnesses were determined in the Cousins $B$ and $R$ bands. Best
fitting exponential model parameters, i.e. extrapolated central
surface brightness and exponential scale length were also
determined. Finally, color gradients were measured for all but two
galaxies lacking images in both colors.  The tables and profile data
in electronic format are available from us on request.

\section{Sample and imaging}
\label{sample}

The photometry of the objects listed in Table \ref{cnvsample} is based
on images taken during a 14-night run in March 1998 on the
$1.2\:\mathrm{ m}$ telescope of the Observatoire de Haute-Provence
(OHP). Unfortunately, half of the nights were lost to bad weather, so
the total amount of galaxies successfully imaged was 15. The frames
are 40 minute $B$ and 20 minute $R$ exposures. We used the $\mathrm{
n}^{\circ}2$ Tektronix $1024 \times 1024$ CCD camera. The pixel scale
is $0\farcs69^{-1}$, giving a frame size of $11\farcm8 \times
11\farcm8$.  The gain was set to $3.5\,\mathrm {e^-}$ per ADU, and the
CCD was read out in the fast mode, with a readout noise of $8.5
\,\mathrm {e^-}$. Seeing was generally fair to poor, between $2\farcs5$
and $4\farcs5$, but for our purposes the images were still useful. The
data collected in Table \ref{cnvsample} consists of the identification
of the imaged galaxies in columns 2 and 3, their 2000.0 epoch
coordinates in cols.\ 4 and 5, types in col.\ 6, with classification
from Binggeli et al.~\cite*{1990A&A...228...42B} marked with $^b$,
total B magnitude from the present photometry, except for galaxy \#13, 
in col.\ 7, heliocentric radial velocity in col.\ 8 and photometric
distances in the last column. All data except for those of the last
column, the magnitudes and type data marked with $^b$ are taken from
the NED. The photometric distances are from Makarova et al.\
\cite*{1998A&AS..128..459M,1997AstL...23..378M}$^{1,\,4}$, Tikhonov \&
Karachentsev \cite*{1998A&AS..128..325T}$^2$ and Georgiev et al.\
\cite*{1997AstL...23..514G}$^3$.
\begin{table*}
\caption[]{CVnI galaxies observed. See text for explanations.}
\begin{center}
\begin{tabular}{rllD{h}{\mbox{h}}{10}D{d}{\degr}{10}lD{.}{.}{4}cD{.}{.}{2}}
\noalign{\smallskip}
\hline
\noalign{\smallskip}
No. &Ident. 1 & Ident. 2 & \multicolumn{1}{c}{\mbox{R.A.}} & \multicolumn{1}{c}{\mbox{Dec.}} 
& Type & \multicolumn{1}{c}{\mbox{$B_{\mathrm T}$} }
& $V_{\mathrm {hel}}$
&\multicolumn{1}{l}{\mbox{Dist.}}\\
\noalign{\smallskip}
\hline
\noalign{\smallskip}
 1. &\object{UGC 06541}&\object{UGC 06541} &11h33\arcmin29\farcs12 & +49d14\arcmin17\farcs4&SmIII/BCD$^b$   &14.32 &249  &  3.5\mbox{$^3$}\\
 2. &\object{NGC 3738} &\object{UGC 06565} &11h35\arcmin48\farcs47 & +54d31\arcmin27\farcs9&Irr             &11.92 &229  &  3.5\mbox{$^3$}\\
 3. &\object{NGC 3741} &\object{UGC 06572} &11h36\arcmin06\farcs18 & +45d17\arcmin01\farcs1&ImIII/BCD$^b$   &14.38 &229  &  3.5\mbox{$^3$}\\
 4. &\object{DDO 99}   &\object{UGC 06817} &11h50\arcmin52\farcs99 & +38d52\arcmin49\farcs0&Im              &13.45 &243  &  3.9\mbox{$^3$}\\
 5. &\object{NGC 4068} &\object{UGC 07047} &12h04\arcmin02\farcs39 & +52d35\arcmin24\farcs1&SmIII/BCD$^b$   &12.93 &210  &  5.25\mbox{$^4$}\\
 6. &\object{NGC 4150} &\object{UGC 07165} &12h10\arcmin33\farcs36 & +30d24\arcmin11\farcs8&SA(r)           &12.41 &226  &       \\
 7. &\object{NGC 4163} &\object{UGC 07199} &12h12\arcmin09\farcs15 & +36d10\arcmin09\farcs1&BCD$^b$         &13.66 &165  &  3.6\mbox{$^2$}\\
 8. &\object{DDO 113 } &\object{UGCA 276}  &12h14\arcmin57\farcs92 & +36d13\arcmin07\farcs8&dE3$^b$         &15.70 &284  &  4.1\mbox{$^4$}\\
 9. &\object{UGC 07298}&\object{UGC 07298} &12h16\arcmin28         & +52d15\farcm3       &ImIV?$^b$         &15.96 &172  &  8.6\mbox{$^2$}\\
10. &\object{NGC 4248} &\object{UGC 07335} &12h17\arcmin50\farcs36& +47d24\arcmin30\farcs6&d:S0$^b$         &13.13 &484  &       \\
11. &\object{UGC 07356}&\object{UGC 07356} &12h19\arcmin09\farcs6  & +47d05\arcmin29\arcsec  &dE4,N$^b$     &15.58 &272  & \multicolumn{1}{r}{\mbox{unr$^1$}}\\
12. &\object{IC 779  } &\object{UGC 07369} &12h19\arcmin38\farcs76 & +29d53\arcmin00\farcs2&E               &14.84 &333  &       \\
13. &\object{DDO 126 } &\object{UGC 07559} &12h27\arcmin05\farcs15 & +37d08\arcmin33\farcs3&Sdm$^b$         &14.2  &218  &  5.1\mbox{$^1$}\\
14. &\object{DDO 127 } &\object{UGC 07599} &12h28\arcmin28\farcs56 & +37d14\arcmin01\farcs1&Sm              &14.80 &278  &  6.9\mbox{$^1$}\\
15. &\object{UGC 07639}&\object{UGC 07639} &12h29\arcmin53\farcs04& +47d31\arcmin47\farcs9&dS0/BCD$^b$      &13.94 &382  &  8.0\mbox{$^1$}\\
\noalign{\smallskip}
\hline
\noalign{\smallskip}
\end{tabular}
\end{center}
\normalsize
\label{cnvsample}
\end{table*}
A gallery of our images is shown in Fig.~\ref{images}.
\setcounter{figure}{1}
\begin{figure*}[!ht]
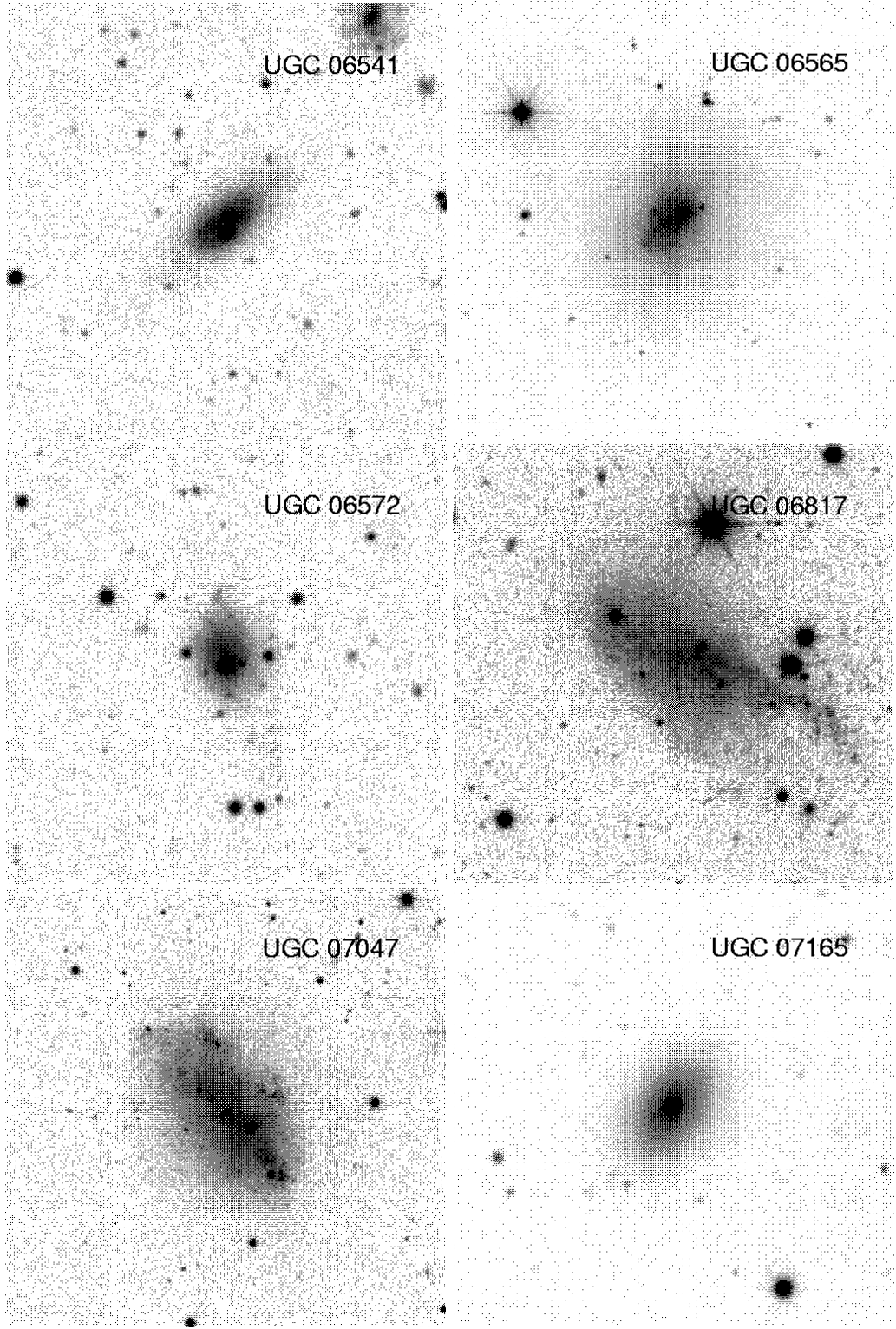

\vspace{0cm}
\hspace{0cm}
\hbox to \textwidth{\hfil
\vbox{\vfil
\hbox{\resizebox{6.75cm}{!}{\includegraphics{smallUGC6541B.eps.small}}
\resizebox{6.75cm}{!}{\includegraphics{small2NGC3738B_A.eps.small}}}
\hbox{\resizebox{6.75cm}{!}{\includegraphics{smallNGC3741B.eps.small}}
\resizebox{6.75cm}{!}{\includegraphics{smallDDO99B_A.eps.small}}}
\hbox{\resizebox{6.75cm}{!}{\includegraphics{smallNGC4068B_A.eps.small}}
\resizebox{6.75cm}{!}{\includegraphics{small2NGC4150B_A.eps.small}}}
\vfil}
\hfil}
\caption{$B$-band CCD images (except UGC 07559 that was only imaged in
the $R$ band).  The scale is the same for all pictures and is given by
the size of one image side $=5\farcm9$. North is up and east to the
left.}
\label{images}
\end{figure*}
\setcounter{figure}{1}
\begin{figure*}[!ht]
\label{images-b}
\vspace{0cm}
\hspace{0cm}
\hbox to \textwidth{\hfil
\vbox{\vfil
\hbox{\resizebox{6.75cm}{!}{\includegraphics{smallNGC4163B_A.eps.small}}
\resizebox{6.75cm}{!}{\includegraphics{smallDDO113B_A.eps.small}}}
\hbox{\resizebox{6.75cm}{!}{\includegraphics{smallUGC7298B_A.eps.small}}
\resizebox{6.75cm}{!}{\includegraphics{small2NGC4248B_A.eps.small}}}
\hbox{\resizebox{6.75cm}{!}{\includegraphics{smallUGC7356B_A.eps.small}}
\resizebox{6.75cm}{!}{\includegraphics{smallIC779B_A.eps.small}}}
\vfil}
\hfil}
\caption{continued}
\end{figure*}
\setcounter{figure}{1}
\begin{figure*}[!ht]
\vspace{0cm}
\hspace{0cm}
\hbox to \textwidth{\hfil
\vbox{\vfil 
\hbox{\resizebox{6.75cm}{!}{\includegraphics{smallDDO126R_A.eps.small}}
\resizebox{6.75cm}{!}{\includegraphics{smallDDO127B_A.eps.small}}}
\hbox{\resizebox{6.75cm}{!}{\includegraphics{smallUGC7639B_A.eps.small}}}
\vfil}
\hfil}
\caption{continued}
\label{images-e}
\end{figure*}

\section{Reductions}
\label{reductions}
The photometry was done with the {\tt MIDAS} package developed by {\tt
ESO}. The images were combined, bias-subtracted and flat\-fielded
using standard procedures.  The images were flat-fielded using
combined morning and evening twilight flat-fields as well as dome
flats.  The following reduction steps were done within the {\tt
SURFPHOT} context in {\tt MIDAS}. The background was determined by
fitting a tilted plane with {\tt FIT/BACKGROUND} and was checked for
correctness by measuring the sky level in different locations in the
frames in regions unaffected by the objects. We chose to restrict the
sky fitting to a tilted plane to avoid introducing features which are
difficult to assess, as would be the case if we had used a polynomial
fit. This also stresses the need for precise flat-fielding. The
photometric calibration was done using standard methods. The
calibration fields were chosen to be close on the sky to the observed
galaxies. These were taken from Smith et al.\ \cite*{smith85}. The
calibration stars were imaged before and after imaging every second
galaxy, so one could have a handle on eventual transparency variations
on a time scale of two hours.  For each galaxy the center and the
ellipse-fit parameters, i.e. ellipticity and position angle (counted
counter-clockwise from the horizontal axis), were determined at the
level of $\sim25^{\mathrm{th}}\mathrm{mag}/\sq{\arcsec}$ by the
ellipse fitting routine {\tt FIT/ELL3} and are given in Table
\ref{paandell}.  \newcolumntype{e}{D{.}{.}{2}}
\begin{table}
\caption[]{Ellipse fit parameters at the level of 
$25 \,\mathrm{mag}/\sq{\arcsec}$}
\begin{center}
\begin{tabular}{rleeee}
\noalign{\smallskip}
\hline
\noalign{\smallskip}
No.	&Galaxy	& \multicolumn{1}{c}{\mbox{PA [\degr]}} & \multicolumn{1}{c}{\mbox{a[$\arcsec$]}}&
 \multicolumn{1}{c}{\mbox{b[$\arcsec$]}} &\multicolumn{1}{c}{\mbox{$b/a$ }}\\
\noalign{\smallskip}
\hline
\noalign{\smallskip}
 1.&\object{UGC 06541}&	039& 39.9& 21.2& 0.5 \\
 2.&\object{UGC 06565}&	069& 91.0& 71.6& 0.8 \\
 3.&\object{UGC 06572}&	105& 35.9& 27.8& 0.8 \\
 4.&\object{UGC 06817}&	150& 84.1& 39.2& 0.5 \\
 5.&\object{UGC 07047}&	123& 81.3& 46.4& 0.6 \\
 6.&\object{UGC 07165}&	057& 61.6& 43.2& 0.7 \\
 7.&\object{UGC 07199}&	094& 53.0& 35.2& 0.7 \\
 8.&\object{UGCA 276} &	126& 53.8& 34.1& 0.6 \\
 9.&\object{UGC 07298}&	047& 23.0& 15.3& 0.7 \\
10.&\object{UGC 07335}&	020& 87.6& 37.6& 0.4 \\
11.&\object{UGC 07356}& 062& 25.9& 17.6& 0.7 \\
12.&\object{UGC 07369}&	151& 29.9& 24.2& 0.8 \\
13.&\object{UGC 07559}&	056& 64.5& 38.1& 0.6 \\
14.&\object{UGC 07599}&	038& 43.5& 22.4& 0.5 \\
15.&\object{UGC 07639}&	056& 56.3& 30.2& 0.5 \\
\noalign{\smallskip}
\hline
\end{tabular}
\end{center}
\label{paandell}
\end{table}
These parameters were used to obtain the total light profile,
i.e. growth curve, by integrating the galaxy light in elliptical
aperture of fixed ellipticity and position angle but with increasing
equivalent ($r=\sqrt{ab}$) radius. Bright sources were removed by hand
from the frames before integrating the light. 

\section{Results}
\label{results}
\subsection{Model-free photometric parameters and radial profiles}
\label{photometry}
The global photometric parameters of our objects are listed in Table
\ref{globalprop}, and the columns represent: the galaxies number
ordered by increasing right ascension (1), name of the galaxy (2),
total apparent magnitude in the $B$ band (3), total apparent magnitude
in the $R$ band (4), effective radius in B in arc seconds (5),
effective radius in R in arc seconds (6), effective surface
bright\-ness in B $[\mathrm{mag}/\sq{\arcsec}]$ (7), effective surface
bright\-ness in R $[\mathrm{mag}/\sq{\arcsec}]$ (8) , total $B-R$
color index (9) and galactic absorption in $B$ from the NED (10).

\begin{figure*}
\resizebox{\hsize}{!}{\includegraphics{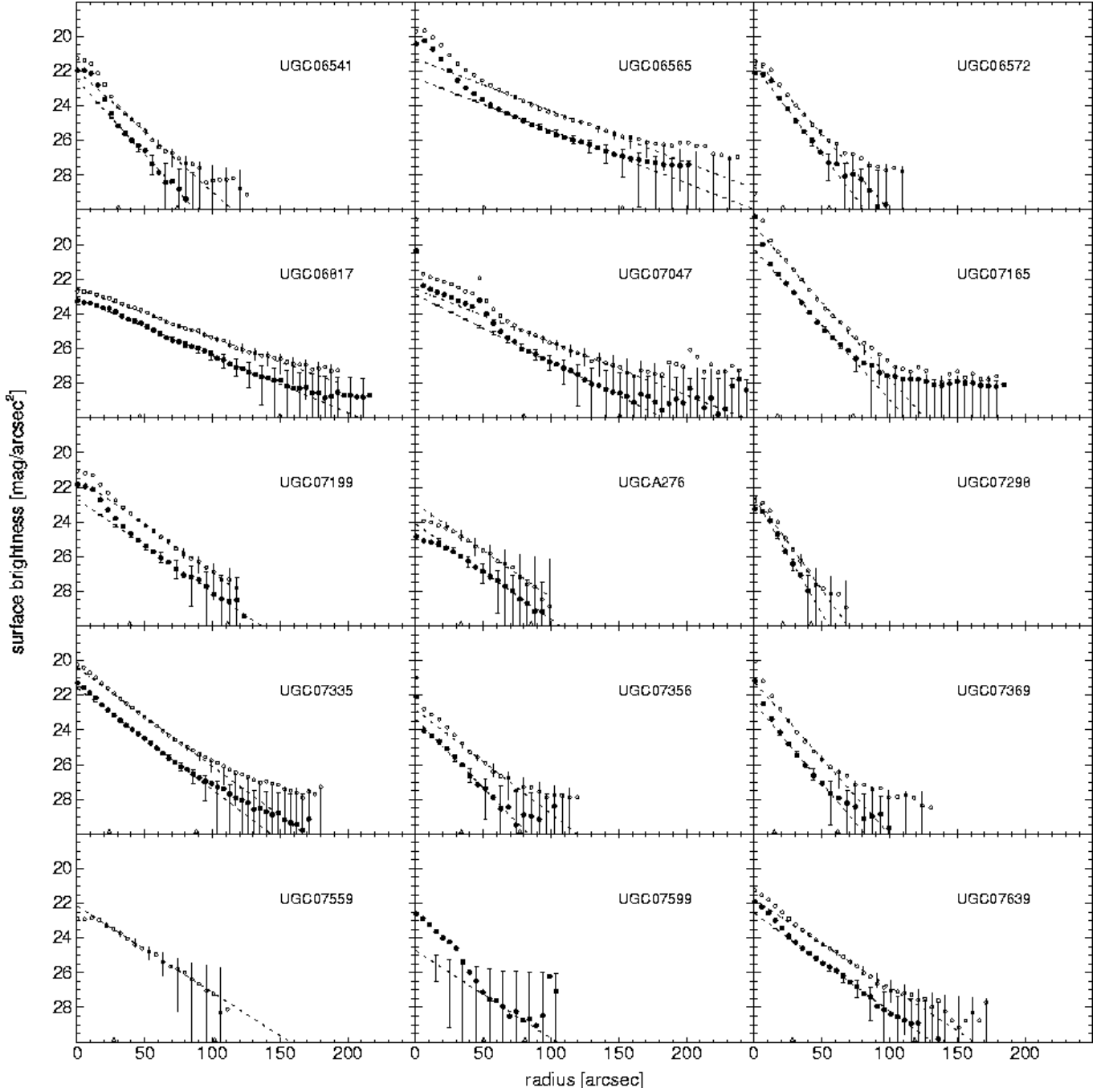}}
\caption{Radial surface brightness profiles of the observed dwarf
galaxies in $B$ (filled circles) and $R$ (open circles) except for UGC
7559 (only $R$) and UGC 7599 (only $B$). The central data point for UGC
7165 in $R$ is at $17.05\,\mathrm{mag}$. The dotted lines represent
the exponential fits, as described in Sect.\ \ref{exponential}, and
the fitting region is marked by triangles on the x-axis. The
error-bars are drawn with hats for the $B$ band data points and without
for the $R$ band data points. These are calculated for (as described in
Sect.~\ref{errors}), and are centered on the best-fitting
exponentials. The radii are all equivalent radii ($r = \sqrt{ab}$). }
\label{profiles}
\end{figure*}

The total apparent magnitude of a galaxy was read off the growth curve
at a sufficiently large radius, i.e. where the growth curve becomes
asymptotically flat. The model-free effective radius was then read at
half of the total growth curve intensity. The effective surface
brightness is then given by
\begin{equation}
\mathopen{<} \mu \mathclose{>}_\mathrm{ eff}[\mathrm{mag}/\sq{\arcsec}]= M + 5
\log(R_\mathrm{ eff}[{\arcsec}]) + 2.
\label{mueffdef}
\end{equation}
All radii refer to equivalent radii, $r=\sqrt{ab}$, where $a$ and $b$
are the major and minor axis of the galaxy, respectively.  Surface
brightness profiles were obtained by differentiating the growth curves
with respect to equivalent radius.  The resulting $B$ and $R$ surface
brightness profiles are shown in Fig.~\ref{profiles}.  The profiles
are traced down to the level where the uncertainties owing to the
fluctuations in the sky level on the profile become dominant.  As
discussed in Sect.\ \ref{errors}, this represents approx.\
$28.5\,\mathrm{mag}/ \sq{\arcsec}$ in $B$ and $27.5\,\mathrm{mag}/
\sq{\arcsec}$ in $R$.  Galactic absorption values were taken from the
NED database. A correction for internal extinction was not applied, as
it is not well known in the case if dwarf galaxies.  The profiles
drawn in the figures have been slightly smoothed with a running window
of width $\approx 5 \arcsec$ and are plotted on a linear radius scale.
\newcolumntype{d}{D{.}{.}{3}}
\begin{table*}
\caption[]{Global photometric properties. See text for explanations.}
\begin{center}
\begin{tabular}{rldddddddd}
\noalign{\smallskip}
\hline
\noalign{\smallskip}
No.	&Galaxy	&\multicolumn{1}{c}{\mbox{$B_{\mathrm T}$}}&\multicolumn{1}{c}{\mbox{$R_{\mathrm T}$}}&
\multicolumn{1}{c}{\mbox{$r^B_\mathrm{ eff}$}}&\multicolumn{1}{c}{\mbox{$r^R_\mathrm{ eff}$}}
&\multicolumn{1}{c}{\mbox{$\mathopen{<} \mu 
\mathclose{>}^B_\mathrm{ eff}$}}&\multicolumn{1}{c}{\mbox{$\mathopen{<} \mu \mathclose{>}^R_\mathrm{ eff}$}}&
\multicolumn{1}{c}{\mbox{$B-R$}}& \multicolumn{1}{c}{\mbox{$A_B$ }}\\
\noalign{\smallskip}
\hline
\noalign{\smallskip}
 1.&\object{UGC 06541}  &14.32 & 13.46 & 14.99 & 17.98&  22.20 & 21.74 &  0.86&0.00 \\
 2.&\object{UGC 06565}  &11.92 & 10.94 & 28.60 & 37.12&  21.20 & 20.79 &  0.98&0.00 \\
 3.&\object{UGC 06572}  &14.38 & 13.57 & 16.30 & 19.92&  22.44 & 22.07 &  0.81&0.00 \\
 4.&\object{UGC 06817}  &13.45 & 12.61 & 53.41 & 61.37&  24.09 & 23.55 &  0.85&0.00 \\
 5.&\object{UGC 07047}  &12.93 & 11.93 & 42.00 & 46.14&  23.05 & 22.25 &  1.00&0.00 \\
 6.&\object{UGC 07165}  &12.41 & 11.11 & 16.09 & 15.52&  20.44 & 19.06 &  1.30&0.04 \\
 7.&\object{UGC 07199}  &13.66 & 12.65 & 22.92 & 27.39&  22.46 & 21.83 &  1.02&0.00 \\
 8.&\object{UGCA 276}   &15.70 & 14.58 & 38.24 & 37.69&  25.62 & 24.46 &  1.13&0.00 \\
 9.&\object{UGC 07298}  &15.96 & 15.25 & 13.46 & 15.70&  23.61 & 23.23 &  0.72&0.04 \\
10.&\object{UGC 07335}  &13.13 & 11.87 & 29.66 & 31.01&  22.49 & 21.33 &  1.26&0.00 \\
11.&\object{UGC 07356}  &15.58 & 14.21 & 23.78 & 26.61&  24.46 & 23.33 &  1.37&0.00 \\
12.&\object{UGC 07369}  &14.84 & 13.46 & 16.07 & 17.30&  22.87 & 21.65 &  1.38&0.07 \\
13.&\object{UGC 07559}  &      & 13.55 &       & 35.86&        & 23.32 &      &0.00 \\
14.&\object{UGC 07599}  &14.80 &       & 22.15 &      &  23.52 &       &      &0.00 \\
15.&\object{UGC 07639}  &13.94 & 12.93 & 26.11 & 31.64&  23.02 & 22.43 &  1.01&0.00 \\
\noalign{\smallskip}
\hline
\end{tabular}
\end{center}
\label{globalprop}
\end{table*}

\begin{figure*}
\resizebox{\hsize}{!}{\includegraphics{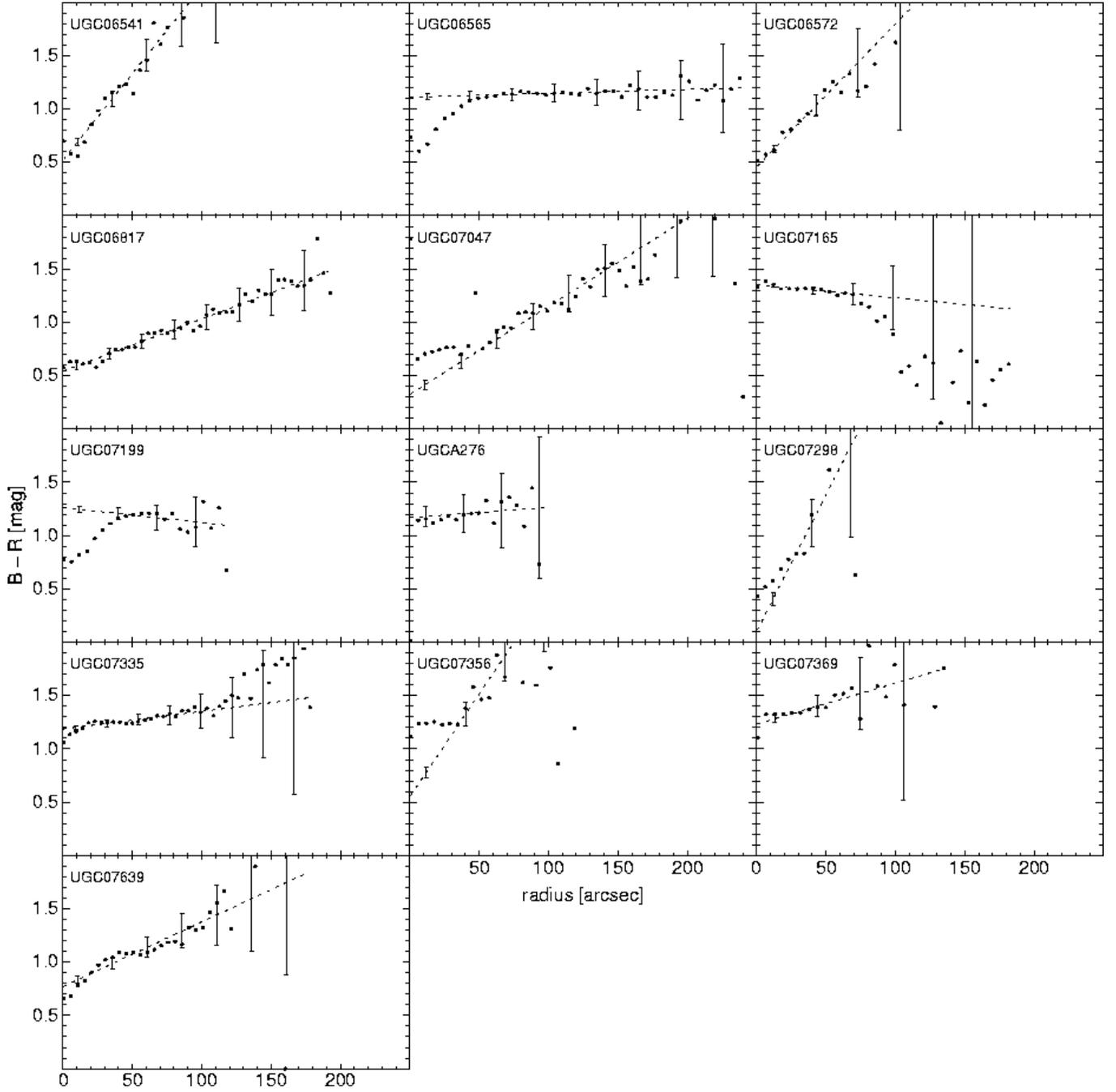}}
\caption{Radial $B-R$ color profiles. The dotted lines represent the
exponential fits, as described in Sect.~\ref{exponential}. The
error-bars have been computed as described in Sect.~\ref{errors}. }
\label{colourprof}
\end{figure*}

Color profiles obtained by subtracting the $R$ from the $B$ surface
brightness profiles, together with the difference between the slopes of
exponential fits (dotted line, see \ref{exponential}) are plotted in
Fig.~\ref{colourprof}.

\subsection{The exponential model: fits and parameters}
\label{exponential}
It is well accepted that the radial intensity profiles of dwarf
galaxies can be reasonably well fitted by a simple exponential
\cite{deV59,BC93}.  This applies not only for dwarf ellipticals, but
also for irregulars, if one looks aside from the brighter star-forming
regions and considers the underlying older populations.  These
profiles can be written as
\begin{equation}
I(r) = I_0 \:\exp{\left(-{r\over r_0}\right)}\equiv I_0\:e^{-\alpha r},
\label{exp}
\end{equation}
that in surface bright\-ness representation becomes a straight line
\begin{equation}
\mu (r)= \mu_0 + 1.086\: \alpha r.
\label{expdr}
\end{equation}
The central extrapolated surface bright\-ness $\mu_0$ and the
exponential scale length $1/\alpha$ are the two free parameters of the
exponential fit.  In this work the fits to the profiles were done on
the outer parts of the profiles by a least squares fitting procedure.
The best-fitting parameters are listed in Table~\ref{modelparam}. The
best-fitting exponential profiles are plotted as dash-dotted lines
along with the observed profiles in Fig.~\ref{profiles}.  The
deviation from a pure exponential law is expressed by the difference
 $\Delta M$ between the total magnitude of an exponential intensity law given by
\begin{equation}
M_{\exp}=\mu_0^{\exp} + 5\log\alpha -2.0,
\label{Mexp}
\end{equation}
and the actual measured total magnitude.  The results are shown in
Table~\ref{modelparam}.  The difference shows the goodness of fit of
the exponential intensity profile.  The columns of
Table~\ref{modelparam} are as follows: (1) as column 1 of Table
\ref{globalprop}, (2) as column 2 of Table \ref{globalprop},
extrapolated central surface bright\-ness according to equation
\ref{expdr} in $B \,[\mathrm{mag}/\sq{\arcsec}]$ (3), same in $R$ (4),
exponential scale length in $B\,[\arcsec]$ (5), same in $R$ (6),
difference between the total magnitude as derived from the exponential
model and the true total magnitude in $B$ (7), same in $R$ (8), radial
color gradient determined from the difference in the slopes of the
model fits as described in Sect.~\ref{results}
$[\mathrm{mag}/\arcsec]$ (9).  \newcolumntype{d}{D{.}{.}{3}}
\begin{table*}
\caption[]{Model parameters. See text for explanations}
\begin{center}
\begin{tabular}{rlddddddd}
\noalign{\smallskip}
\hline
\noalign{\smallskip}
No.	&Galaxy	&\multicolumn{1}{c}{\mbox{$(\mu^{\exp}_0)_B$}}&\multicolumn{1}{c}{\mbox{$(\mu^{\exp}_0)_R$}}&
\multicolumn{1}{c}{\mbox{$1/{\alpha_B}$}}&\multicolumn{1}{c}{\mbox{$1/{\alpha_R}$
}}&\multicolumn{1}{c}{\mbox{$\Delta M_B$}}&\multicolumn{1}{c}{\mbox{$\Delta M_R$}}&
\multicolumn{1}{c}{\mbox{$\mathrm{ d}(B-R)\over \mathrm{ d}r$}}\\ 
\noalign{\smallskip}
\hline
\noalign{\smallskip}
 1.&\object{UGC 06541}  &22.49 &  21.97  & 12.60  & 15.53  &  0.67  &  0.55  &  0.016 \\
 2.&\object{UGC 06565}  &22.40 &  21.29  & 35.88  & 36.31  &  0.71  &  0.55  &  0.000 \\
 3.&\object{UGC 06572}  &21.71 &  21.25  & 10.73  & 12.37  &  0.18  &  0.22  &  0.013 \\
 4.&\object{UGC 06817}  &23.09 &  22.55  & 32.83  & 38.56  &  0.05  &  0.01  &  0.005 \\
 5.&\object{UGC 07047}  &22.89 &  22.57  & 27.93  & 35.46  &  0.73  &  0.89  &  0.008 \\
 6.&\object{UGC 07165}  &20.29 &  18.94  & 12.54  & 12.36  &  0.39  &  0.37  & -0.001 \\
 7.&\object{UGC 07199}  &22.68 &  21.41  & 20.23  & 19.71  &  0.48  &  0.29  & -0.001 \\
 8.&\object{UGCA 276}   &24.12 &  22.95  & 19.75  & 20.10  & -0.06  & -0.14  &  0.001 \\
 9.&\object{UGC 07298}  &22.54 &  22.44  &  8.05  &  9.93  &  0.05  &  0.21  &  0.026 \\
10.&\object{UGC 07335}  &21.57 &  20.38  & 18.48  & 19.01  &  0.10  &  0.12  &  0.002 \\
11.&\object{UGC 07356}  &23.37 &  22.82  & 13.65  & 18.03  &  0.11  &  0.33  &  0.019 \\
12.&\object{UGC 07369}  &22.47 &  21.24  & 11.82  & 12.34  &  0.27  &  0.32  &  0.004 \\
13.&\object{UGC 07559}  &      &  22.14  &        & 21.75  &        & -0.09  &        \\
14.&\object{UGC 07599}  &24.72 &         & 21.41  &        &  1.27  &        &        \\
15.&\object{UGC 07639}  &22.57 &  21.79  & 19.24  & 21.56  &  0.21  &  0.20  &  0.006 \\
\noalign{\smallskip}
\hline
\end{tabular}
\end{center}
\label{modelparam}
\end{table*}

%
\subsection{Photometric uncertainties}
\label{errors}
The photometric calibration represents the greatest source of error on
the global parameters. The nights were not photometric, so
calibrations were done with fields close on the sky to the actual
objects, see section \ref{reductions}. Some nights were rejected
altogether because of the changing transparency on short
time scales. The combined (zero point and slope of the extinction curve)
statistical uncertainty on the photometric calibration is estimated to
be about $0.1$ mag.  The uncertainties on the surface brightness
profiles (SBP) are a combination of Poisson noise and, at low levels,
the large-scale sky fluctuations, owing to flat fielding and sky
subtraction imprecision. The uncertainties shown in Fig.\
\ref{profiles} have been calculated assuming the galaxies have pure
exponential SBPs. To the Poisson noise caused by the photons we added
a constant contribution corresponding to $0.5\%$ of the actual
measured sky background, that combines the flat field and sky
subtraction terms mentioned above. This level of accuracy for the
flat fielding and sky subtraction was conservatively estimated by
inspecting many frames. Most frames show better flat fielding accuracy.
The uncertainties on the profiles were then calculated for azimuthally
averaged annuli of one arc second width\footnote{the profiles in Fig.\
\ref{profiles} have been smoothed for plotting, so the error-bars apply
to the original un-smoothed SBPs. Also, the calibration uncertainties
are not shown in the plots.}. For the actual calculations, we
considered circles instead of ellipses and compared equivalent radii
when determining the error at a given point on a SBP. The galaxy
surface brightness and the fluctuations in the sky background reach
similar levels at $\sim 28.5$ in $B$ and $\sim 27.5
\,\mathrm{mag}/\sq{\arcsec}$ in $R$.
 
The errors on the color profiles, shown in Fig.~\ref{colourprof},
were estimated using the error term as described above for each color
and applying usual error formulae for logarithms and combining the
errors thus obtained for each color by quadrature.

\begin{figure}
\resizebox{\columnwidth}{!}{\includegraphics{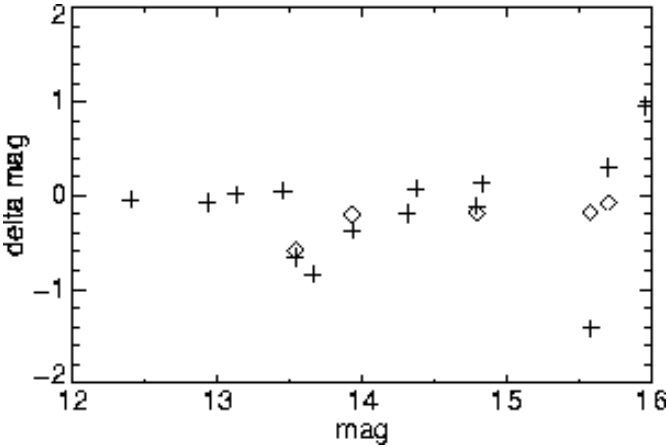}}
\caption{Comparison between the photometry of the present paper and
data from the RC3 catalog (crosses) and from Makarova et
al. (lozenges). The $B$ magnitudes are from the present paper, and delta
mag is the difference between our photometry and RC3 or Makarova et
al.\ (1997, 1998)}
\label{compRC3}
\end{figure}

Comparing the $B$ magnitudes from this paper with RC3 data from the
NED, one sees that our photometry agrees well with the RC3 for ten
galaxies, see Fig.~\ref{compRC3}. Three galaxies, namely UGC 07199,
UGC 07356 and UGC 07559 have brighter magnitudes and one, UGC 7298,
has a fainter magnitude in this paper as compared to the values in the
RC3. For a subsample of our galaxies, there are total magnitudes
calculated by Makarova et al.\
\cite*{1997AstL...23..378M,1998A&AS..128..459M}. Their data is shown
in Fig.\ \ref{compRC3} as lozenges. The agreement between both
photometries is better than for the RC3 data, especially for UGC 7356.

\section{Notes}
 The galaxies presented here have a median absolute magnitude of
$-14.43\,\mathrm{mag}$. There are no very faint dwarf galaxies in this
sample (see Fig. \ref{compRC3}, compare with Paper I). The lowest
magnitude object, UGCA 276, is at $-12.37\,\mathrm{mag}$, and is
classified as a dwarf elliptical.
\begin{figure}
\resizebox{8cm}{!}{\includegraphics{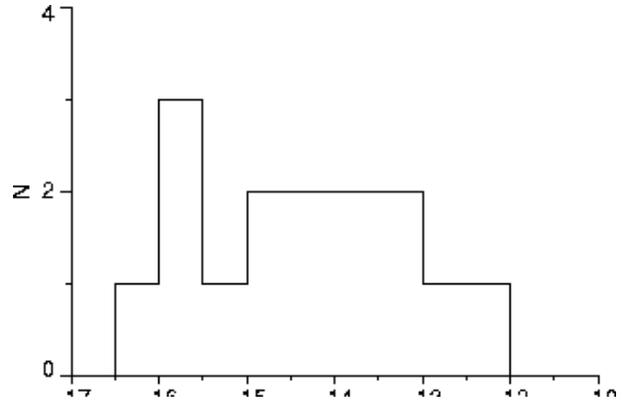}}
\vspace{2mm}
\caption{Distribution of galaxy magnitudes. Distances are from Table
\ref{cnvsample}. Where photometric distances were unavailable, the
radial velocities were used with $H_0=60\:\mathrm{km\,s^{-1}\,Mpc^{-1}}$}
\label{absm}
\end{figure}
%
%
%

\noindent\emph{On individual galaxies}:\\
\noindent {\bfseries \object{UGC 6541}}: small galaxy, with a light
excess close to the center as compared to an exponential profile. As
can be seen on the image, this can be a star-forming region or a
superimposed star. The color profile shows no deviation in the region
concerned by the bright object, so its nature is difficult to
assess. The color gradient of this galaxy is steep, and in the sense
of the center being bluer.

\noindent {\bfseries \object{UGC 6565}}: possesses a bright central
light excess above a pure exponential. This central part also has a
knotty structure, and is also much bluer than the rest of the
galaxy. This probably suggests ongoing star formation. The outer
part of the galaxy seems smooth. The isophotes look boxy, although we
haven't quantified this feature in this work.

\noindent {\bfseries \object{UGC 6572}}: irregular structure, with a
slightly offset bright region. The color profile is steep, but the
brighter spot doesn't seem to be of a different color than its
immediate surroundings.

\noindent {\bfseries \object{UGC 6817}}: has a low central surface
brightness and a large scale length. The SBP is close to a pure
exponential, and the color gradient is smooth, showing a reddening
with increasing radius. The aspect of the galaxy is peculiar, with a
knotty irregular surface, and tail extending to the south-west. It
could have been disrupted by supernova events. The isophotes also
seem twisted, although not examined in detail in this work.

\noindent {\bfseries \object{UGC 7047}}: irregular with a knotty
structure, and a light excess over an exponential profile towards the
center. This part of the galaxy also shows a flatter color index,
whereas the outer parts of the galaxy tend to get redder with
increasing radius.

\noindent {\bfseries \object{UGC 7165}}: has a SBP that is close to an
exponential, and also has a bright nuclear part. The color is almost
constant over the whole radius range, apparently including the central
part. But the resolution of our data is insufficient to discuss the
center in any detail. The galaxy as a whole has a smooth elliptical
appearance.

\noindent {\bfseries \object{UGC 7199}}: knotty irregular with a
central blue light excess. Exponential outer SBP shape, with a flat
color profile outside the blue center.

\noindent {\bfseries \object{UGCA 276}}: small galaxy with a faint
central surface brightness. The central part is slightly deficient
compared to an exponential. The color profile is flat.

\noindent {\bfseries \object{UGC 7298}}: apparently small irregular
galaxy with a knotty surface. Its color profile seems steep, being
redder at large radius, but suffers from poor resolution and large
error bars.

\noindent {\bfseries \object{UGC 7335}}: elongated irregular galaxy
with a bright central lane. The shape of the galaxy as a whole is
elongated and elliptical. Its SBP is close to a pure exponential. The
color profile is shallow, getting slightly redder with radius, and
showing a bluing close to the center.

\noindent {\bfseries \object{UGC 7356}}: classical
nucleated dwarf elliptical. It has a published heliocentric radial
velocity suggesting it is close-by (but it is close in projection to
NGC 4258). The brightest stars were not resolved by Makarova et al.\
\cite*{1998A&AS..128..459M}, confirming the dwarf elliptical
nature. Its color profile is flat, with a slightly bluer nucleus.

\noindent {\bfseries \object{UGC 7369}}: is a smooth, elliptically
shaped galaxy, with a bright nucleus. It has a shallow color profile,
becoming redder with radius, and a slightly blue nucleus. It is
classified in the RC3 as an elliptical, but at $v_{\mathrm hel}=333
\mathrm{km\,s^{-1}}$ it could be a dwarf elliptical. Its SBP is close
to exponential.

\noindent {\bfseries \object{UGC 7559}}: irregular with a knotty
face. Also seems disrupted, although not as severely as UGC 6817. We
only have a R frame. SBP close to an exponential, with a flat central
part.

\noindent {\bfseries \object{UGC 7599}}: apparently small irregular
with a large central light excess above an exponential fit to the
outer regions. Only a B frame was available.

\noindent {\bfseries \object{UGC 7639}}: irregular with a slight
excess above an exponential in the inner part of its SBP. Knotty inner
surface, and a slight luminosity excess to the north-west.  Smooth
color profile, showing a reddening with increasing radius.

Many authors report that color profiles show small gradients or are
flat in dwarf galaxies \cite{1998A&AS..129..313B,patthuan96}.  We here
find that all but two galaxies with color profiles show a reddening
with increasing radius. UGC 6565 and UGC 7199 have a blue central
part when compared to the outer parts. UGC 7047, UGC 7356
and UGC 7298 have flatter color gradients towards their centers. The
remaining galaxies show monotonous color profiles.

To summarize, we have presented global photometric parameters for 15
galaxies in the Canes Venatici Cloud I in the $B$ and $R$ bands. Our
sample corresponds to approx.\ one third of all galaxies in the
direction of the cloud with heliocentric velocities less than $500\:
\mathrm{km\,s^{-1}}$.


\begin{acknowledgements}
T.\ B.\ and B.\ B.\ thank the Swiss National Science Foundation for
financial support.  We also thank Bernhard Parodi for taking part in
the observing run.

This research has made use of the NASA/IPAC Extragalactic Database
(NED) which is operated by the Jet Propulsion Laboratory, California
Institute of Technology, under contract with the National Aeronautics
and Space Administration, as well as  NASA's Astrophysics Data System
Abstract Service. 

\end{acknowledgements}

\bibliographystyle{astron}
\bibliography{mnemonic,canesv}

\end{document}